\documentclass[aps,twocolumn,prd,nofootinbib,showpacs]{revtex4}

\usepackage{amsmath}
\usepackage{amssymb}
\usepackage{graphicx}

\newcommand{\s}{\sigma}
\newcommand{\vb}{\bar{v}}
\newcommand{\tb}{\bar{t}}
\def\bdi#1{\ensuremath{\boldsymbol{#1}}}

\def\bX{{\bdi X}}

\def\bk{{\bdi k}}

\def\Xd{\dot X}

\def\Xp{\acute X}

\def\bXd{\dot\bX}
\def\bXp{\acute\bX}

\newcommand{\de}{\delta}
\newcommand{\ep}{\varepsilon}
\newcommand{\ka}{\kappa}
\newcommand{\si}{\sigma}
\newcommand{\Ga}{\Gamma}
\newcommand{\Om}{\Omega}
\newcommand{\half}{\frac{1}{2}}
\newcommand{\vev}[1]{\langle#1 \rangle}
\newcommand{\lvev}[1]{\left\langle#1 \right\rangle}

\newcommand{\deTovT}{\Theta}
\newcommand{\Area}{\mathcal{A}}

\newcommand{\sss}[1]{{\scriptscriptstyle{#1}}}
\newcommand{\vect}[1]{\boldsymbol{#1}}
\newcommand{\uCMB}{\mathrm{\sss{CMB}}}

\newcommand{\un}{\mathrm{n}}

\newcommand{\uc}{\mathrm{c}}
\newcommand{\us}{\mathrm{s}}

\newcommand{\uuc}{\mathrm{uc}}
\newcommand{\ud}{\mathrm{d}}
\newcommand{\up}{\mathrm{p}}

\newcommand{\Tcmb}{T_\uCMB}
\newcommand{\tension}{U}      

\newcommand{\calD}{\mathcal{D}}
\newcommand{\calC}{\mathcal{C}}
\newcommand{\calJ}{\mathcal{J}}
\newcommand{\corr}{\hat{\xi}}
\newcommand{\alphac}{c_1}
\newcommand{\be}{\vect{e}}
\newcommand{\why}{Y}
\newcommand{\betac}{c_2}
\newcommand{\chimax}{\Lambda}
\newcommand{\gammain}[2]{\gamma_\un\negthinspace\left(#1,#2\right)}
\newcommand{\NL}{\mathrm{NL}}
\newcommand{\tauNL}{\tau_\NL}
\newcommand{\gNL}{g_\NL}

\begin{document}

\title{CMB temperature trispectrum of cosmic strings}

\author{Mark Hindmarsh}
\email{m.b.hindmarsh@sussex.ac.uk}
\affiliation{Department of Physics \& Astronomy, University of Sussex,
 Brighton, BN19QH, United Kingdom}

\author{Christophe Ringeval}
\email{christophe.ringeval@uclouvain.be}
\affiliation{Theoretical and Mathematical Physics Group, Centre for
 Particle Physics and Phenomenology, Louvain University, 2 Chemin du
 Cyclotron, 1348 Louvain-la-Neuve, Belgium}

\author{Teruaki Suyama}
\email{teruaki.suyama@uclouvain.be}
\affiliation{Theoretical and Mathematical Physics Group, Centre for
 Particle Physics and Phenomenology, Louvain University, 2 Chemin du
 Cyclotron, 1348 Louvain-la-Neuve, Belgium}

\date{\today}

\begin{abstract}
  We provide an analytical expression for the trispectrum of the
  Cosmic Microwave Background (CMB) temperature anisotropies induced
  by cosmic strings. Our result is derived for the small angular
  scales under the assumption that the temperature anisotropy is
  induced by the Gott--Kaiser--Stebbins effect. The trispectrum is
  predicted to decay with a non-integer power-law exponent
  $\ell^{-\rho}$ with $6<\rho<7$, depending on the string
  microstructure, and thus on the string model. For Nambu--Goto
  strings, this exponent is related to the string mean square velocity
  and the loop distribution function. We then explore two classes of
  wavenumber configuration in Fourier space, the kite and trapezium
  quadrilaterals. The trispectrum can be of any sign and appears to be
  strongly enhanced for all squeezed quadrilaterals.
\end{abstract}
\pacs{98.80.Cq, 98.70.Vc}
\maketitle

\section{Introduction}

Although cosmic strings may be of various early universe
origins~\cite{Kibble:1976sj,Dabholkar:1990yf, Hindmarsh:1994re,
  VilShe94, Kofman:1994rk,Yokoyama:1989pa, Sakellariadou:2006qs,
  Copeland:2003bj, Sarangi:2002yt, Dvali:2003zj}, being line-like
gravitational objects, they induce temperature discontinuities in the
CMB through the Gott--Kaiser--Stebbins (GKS) effect
\cite{Gott:1984ef,Kaiser:1984iv}. Direct searches for such
discontinuities have been performed without success but do provide
upper limits to the string tension $U$~\cite{Jeong:2004ut, Jeong:2007,
  Christiansen:2008vi}. On the other hand, if cosmic strings are added
to the standard power-law $\Lambda$CDM model~\cite{Bouchet:2000hd}, it
has been shown in Refs.~\cite{Battye:2006pk,Bevis:2007gh} that the CMB
data are fitted even better if the fraction of the temperature power
spectrum due to strings is about $10\%$ (at $\ell=10$). Such a
fraction of string would even dominate the primary anisotropies of
inflationary origin for $\ell \gtrsim 3000$~\cite{Fraisse:2007nu}.
With the advent of the arc-minute resolution CMB experiments and the
soon incoming Planck satellite data, it is therefore crucial to
develop reliable tests for strings~\cite{Seljak:2006}, as to
understand the non-Gaussian signals. The probability distribution of
the fluctuations due to the GKS effect is known to be skewed and has a
less steep decay than Gaussian~\cite{Fraisse:2007nu}, a feature which
can be explained in a simple model of kinked
string~\cite{Takahashi:2008ui}. In Ref.~\cite{Hindmarsh:2009qk}, we
have studied the temperature bispectrum induced by cosmic strings both
analytically and numerically by using Nambu--Goto string
simulations. We found good agreement between the analytical and
numerical bispectrum for both the overall amplitude and the
geometrical factor associated with various triangle configurations of
the wavevectors. This agreement suggests that our analytical
assumptions are capturing the relevant non-Gaussian features of a
string network and could be used to derive other statistical
properties. In this paper, we present new results concerning the
trispectrum, i.e. the four-point function of the temperature
anisotropy~\cite{Hu:2001fa}. As pointed out in
Ref.~\cite{Hindmarsh:2009qk}, the bispectrum is generated only when
the background spacetime breaks the time reversal symmetry. Because
our universe is expanding, the time reversal symmetry is indeed broken
and we get a non-vanishing string bispectrum. On the other hand, the
trispectrum can be generated even in Minkowski spacetime and one may
naively expect a stronger non-Gaussian signal than for the
bispectrum. Motivated by this observation, we provide in this paper an
analytical derivation of the string trispectrum and study its
dependency for various quadrilateral configurations in Fourier space.
Given the fact that analytical predictions and numerical results
exhibit good agreement both for the power
spectrum~\cite{Hindmarsh:1993pu, Fraisse:2007nu} and the
bispectrum~\cite{Hindmarsh:2009qk}, we expect our result here to agree
as well with the numerics. Performing such a comparison would however
require a significant amount of computing resources which motivated us
to leave it for a future work. Our main result can be summarized by
Eq.~(\ref{eq:finaltri}). Interestingly, the power-law behaviour of the
trispectrum exhibits a non-integer exponent which can be related to
the small scale behaviour of the string tangent vector correlator. In
the framework of Nambu--Goto strings, this exponent is related to the
mean square string velocity~\cite{Polchinski:2006ee, Copeland:2009dk}
and to the scaling loop distribution function~\cite{Ringeval:2005kr,
  Rocha:2007ni}.  The paper is organised as follows. In the next
section we briefly recall the assumptions at the basis of our
analytical approach and derive the trispectrum in
Sec.~\ref{sec:trispectrum}. As an illustration, we apply our result to
some specific quadrilaterals in Sec.~\ref{sec:quad} and exhibit some
configurations that leads to a divergent trispectrum. They should
provide the cleanest way to look for a non-Gaussian string signal.

\section{Temperature Anisotropy From Cosmic Strings}

\label{sec:temp}

In this section, we briefly review the general basics needed to
calculate $N$-point function of $\Theta \equiv \Delta T/\Tcmb$ at small
angular scales. To study the correlation functions in the small angle
limit, it is enough to consider $\Theta$ on the small patch of the
sky. Then we can approximate this patch as two-dimensional Euclidean
space, which simplifies the calculations. In this limit, the integrated
Sachs--Wolfe effect generated by cosmic strings yields the temperature
anisotropy in the light-cone gauge~\cite{Hindmarsh:1993pu}
\begin{equation}
  -k^2\deTovT_{\bk} = i\ep k_A  \int \ud\si \Xd^A(\si)
  e^{i\bk\cdot\bX(\si)},
  \label{eFT}
\end{equation}
where we have defined
\begin{equation}
  \ep = 8\pi G\tension,
\end{equation}
and $X^A$ ($A=1,2$) is the two-dimensional string position vector
perpendicular to the line of sight. We implicitly assume a summation
on the repeated indices. It is now clear that the power spectrum,
bispectrum, and higher order correlators can be evaluated in terms of
correlation functions of the string network, as projected onto our
backward light-cone. In order to evaluate the statistical quantities
constructed over $\deTovT_{\bk}$, the correlation functions of $\Xd^A$
and $\Xp^B$ have to be known. However, because $\deTovT_{\bk}$ depends
on $\Xd^A$ and $\Xp^B$ in a non-trivial manner, it is extremely
difficult to derive meaningful consequences for the correlation
functions without imposing additional conditions on the string
correlators. In this paper, as done in Refs.~\cite{Hindmarsh:1993pu,
  Hindmarsh:2009qk} we therefore assume that both $\Xd^A$ and $\Xp^B$
obey Gaussian statistics, and this drastically simplifies our
calculations. All the correlation functions of $\deTovT_{\bk}$ can now
be written in terms of the two-point functions only. Using the same
notation as in Ref.~\cite{Hindmarsh:1993pu}, the two-point functions
of the string correlators are
\begin{equation}
  \label{eq:tangent}
  \begin{aligned}
    \lvev{\Xd^A(\si)\Xd^B (\si')} & = \frac{1}{2} \delta^{AB} V(\si-\si'), \\
    \lvev{\Xd^A(\si) \Xp^B (\si')} & =  \frac{1}{2} \delta^{AB} M (\si-\si'), \\
    \lvev{\Xp^A(\si) \Xp^B (\si')} &= \frac{1}{2} \delta^{AB} T(\si-\si').
  \end{aligned}
\end{equation}
Note that an appearance of a term like $\epsilon^{AB} N(\s-\s')$ in
the mixed correlator $\vev{\Xd^A \Xp^B}$, where $\epsilon^{AB}$ is the
anti-symmetric tensor with $\epsilon^{12}=1$, is forbidden due to the
symmetry. As for the bispectrum, we also introduce the
correlator~\cite{Hindmarsh:2009qk}
\begin{align}
  \Gamma(\si-\si') & \equiv \lvev{\left[
      \vect{X}(\si)-\vect{X}(\si')\right]^2} \\ & = \int_{\si'}^\si
  \ud\si_1\int_{\si'}^\si \ud\si_2 T(\si_1-\si_2).
  \label{e:MixCor}
\end{align}
The leading terms are given by~\cite{Hindmarsh:2009qk}
\begin{equation}
\label{eq:LOcorr}
\begin{aligned}
 V(\si) & \to  \left\{
\begin{array}{cl} 
\vb^2 & \si \to 0 \\ 0 & \si \to \infty
\end{array}
\right. ,\\
\Gamma(\si) & \to  \left\{
\begin{array}{cl}
\tb^2\si^2 & \si \to 0 \\
   \hat\xi\si & \si \to \infty
\end{array}
\right.,
\end{aligned}
\end{equation}
where we have defined 
\begin{eqnarray}
  \hat\xi = \Ga'(\infty), \qquad \vb^2 = \lvev{\bXd^2}, \qquad \tb^2 = \lvev{\bXp^2}.
\end{eqnarray}
The correlation length $\hat\xi$ is the projected correlation length
on the backward light-cone, $\bar t^2$ is the mean square projected
tangent vector (of order unity), $\bar v^2$ is the mean square
projected velocity (again of order unity).

\section{Temperature trispectrum}

\label{sec:trispectrum}

In the flat sky approximation the four-point temperature correlation
function is defined as
\begin{equation}
\begin{aligned}
  \lvev{\deTovT_{\bk_1} \deTovT_{\bk_2}
    \deTovT_{\bk_3} \deTovT_{\bk_4}} & =
  T(\bk_1,\bk_2,\bk_3,\bk_4)(2\pi)^2 \\ & \times \de(\bk_1+\bk_2+\bk_3+\bk_4).
\end{aligned}
\end{equation}
Using Eq.~(\ref{eFT}) and a formal area factor $\Area =
(2\pi)^2\de(0)$, the trispectrum\footnote{Notice that our denomination
  ``trispectrum'' here stands for the four-point function and contains
  the unconnected part. This one is however non-vanishing only for
  parallelogram configurations of the wave vectors.}can be written as
\begin{equation}
\begin{aligned}
\label{eq:tristart}
&  T(\bk_1,\bk_2,\bk_3,\bk_4) = \ep^4 \frac{1}{\Area} \delta_{A
    \bar{A}}\delta_{B \bar{B}} \delta_{C \bar{C} } \delta_{D \bar{D} }
  \dfrac{k^{\bar{A}}_1k^{\bar{B}}_2k^{\bar{C}}_3k^{\bar{D}}_4}
  {k_1^2k_2^2k_3^2k_4^2} \\ & \times \int \ud\si_1\ud\si_2\ud\si_3 \ud\si_4
  \lvev{\Xd^A_1\Xd^B_2\Xd^C_3 \Xd^D_4 e^{i\delta^{ab}\bk_a \cdot \bX_b}},
\end{aligned}
\end{equation}
with $\Xd^A_a = \Xd^A(\si_a)$, $(a,b) \in \{1,2,3,4\}$ and
$\bk_1+\bk_2+\bk_3+\bk_4=0$.  We now assume Gaussian statistics and
define
\begin{align}
\calC^{ABCD} &= \Xd^A_1\Xd^B_2\Xd^C_3\Xd^D_4,\\
\calD & = \delta^{ab}\bk_a \cdot \bX_b.
\end{align}
The ensemble average in Eq.~(\ref{eq:tristart}) can be expressed in
terms of the two-point functions only
\begin{equation}
\begin{aligned}
  \lvev{\calC^{ABCD} e^{i\calD}} &= \left[\vev{\calC^{ABCD}} + 
    \vev{\Xd^A_1 \Xd^B_2} \vev{\Xd^C_3 \calD} + \circlearrowleft
  \right. \\
  & \left. +
    \vev{\Xd^A_1 \calD} \vev{\Xd^B_2 \calD} \vev{\Xd^C_3 \calD}
    \vev{\Xd^D_4 \calD}\right] e^{-\half
    \vev{\calD^2}}, \label{decomposition}
\end{aligned}
\end{equation}
where $\circlearrowleft$ denotes permutations of the labels
$\{1,2,3\}$. Expressing $\bX(\sigma)$ in terms of $\bXp(\sigma)$ makes
clear that all terms but the first involve the mixed correlators
$\vev{\bXd(\s) \bXp (\s')}$. Since they give a small contribution
compared to the others, we will not consider these extra terms in the
following. A more detailed calculation would show that they induce
corrections to the trispectrum scaling relative to the first term as
$1/k$ and $1/k^2$, respectively, and therefore are negligible at small
angular scales. The trispectrum can be approximated as
\begin{equation}
\begin{aligned}
  T(\bk_1,\bk_2,\bk_3,\bk_4) & \simeq \ep^4 \frac{1}{\Area} \delta_{A
    \bar{A}}\delta_{B \bar{B}} \delta_{C \bar{C} } \delta_{D \bar{D} }
  \dfrac{k^{\bar{A}}_1k^{\bar{B}}_2k^{\bar{C}}_3k^{\bar{D}}_4}{k_1^2k_2^2k_3^2k_4^2}
  \\ & \times \int \ud\si_1\ud\si_2\ud\si_3 \ud\si_4 \vev{\calC^{ABCD}}
  e^{-\half \vev{\calD^2}}.
\end{aligned}
\label{eq:trispectrum1}
\end{equation}
In terms of the two-point functions introduced in Sec.~\ref{sec:temp},
\begin{equation}
\label{eq:V2term}
\begin{aligned}
  & \vev{\calC^{ABCD}} = \frac14 \de^{AB}\de^{CD} V(\si_{12}) V(\si_{34})
  \\ & + \frac14 \de^{AC} \de^{DB}V(\si_{31})(\si_{42}) + \frac14
  \de^{AD}\de^{BC} V(\si_{14}) V(\si_{23}),
\end{aligned}
\end{equation}
where $\si_{ab}\equiv\si_a - \si_b$.  As for $\vev{\calD^2}$, replacing
$\bk_4$ with $-\bk_1-\bk_2-\bk_3$, one gets
\begin{equation}
\label{eq:D2def}
\lvev{\calD^2} = \lvev{\left( \bk_1\cdot \bX_{14} + \bk_2\cdot
    \bX_{24} + \bk_3\cdot \bX_{34} \right)^2} ,
\end{equation}
where $\bX_{ab} \equiv \bX_a - \bX_b$. As in
Ref.~\cite{Hindmarsh:2009qk}, one can show that
\begin{equation}
\lvev{\bX_{14}\cdot \bX_{24}} = \dfrac{1}{2} \left[ \Gamma(\si_{14})
  + \Gamma(\si_{24}) - \Gamma(\si_{12}) \right],
\end{equation}
which can be used to transform Eq.~(\ref{eq:D2def}) into a manifestly
symmetric expression
\begin{equation}
\label{eq:D2sym}
\lvev{\calD^2} = \dfrac{1}{2} \sum_{a<b} \kappa_{ab} \Gamma(\si_{ab}),
\end{equation}
with
\begin{equation}
\kappa_{ab} \equiv -\bk_a \cdot \bk_b.
\end{equation}
At this point, plugging this expression into
Eq.~(\ref{eq:trispectrum1}) and performing the integrations along the
lines done for the bispectrum is not possible (see
Ref.~\cite{Hindmarsh:2009qk}). Indeed, since the $\bk_a$ are forming a
quadrilateral, contrary to the bispectrum triangle configurations, all
the $\kappa_{ab}$ cannot be positive thereby preventing some of the
Gaussian integrals to be performed.

We can nevertheless perform one integration by switching to the more
convenient integration variables $\si_{14}$, $\si_{24}$, $\si_{34}$
and $\si_4$. The Jacobian is unity and Eq.~(\ref{eq:D2sym}) can be
rewritten in a non-symmetric form depending only on three of the
variables:
\begin{equation}
\label{eq:D2Omega}
  \lvev{\calD^2} = -\dfrac{1}{2} \sum_{i=1}^3\sum_{j=1}^3 \kappa_{ij}\Omega_{ij},
\end{equation}
where
\begin{equation}
  \Omega_{ij} = \dfrac{1}{2} \left[\Gamma(\si_{i4}) + \Gamma(\si_{j4}) -
    \Gamma(\si_{i4} - \si_{j4}) \right].
\end{equation}
From Eqs.~(\ref{eq:trispectrum1}) and (\ref{eq:V2term}), we find that
the integrand does not depend on $\si_4$ and the integration yields a
factor equal to the total length of the strings $L$ in the area
$\Area$. In order to perform the integration over the other variables,
one can again use the small angle approximation where all the $k_a$
are taken to be sufficiently large.  The dominant parts then come from
the small $\si$ length scales, the contributions from other regions
being exponentially suppressed. This suggests we should Taylor-expand the
two-point functions around $\si=0$. At leading order, using
Eq.~(\ref{eq:LOcorr}), one gets $\Omega_{ij} \simeq \tb^2 \si_{i4}
\si_{j4}$ implying that $\vev{\calD^2}$ is a quadratic form in the
variables $\si_{i4}$. However, it exhibits a vanishing eigenvalue and
the Gaussian integral cannot be extended to infinity since there is
one direction of integration along which the exponent
$\ka_{ij}\Om_{ij}$ remains null. Let us notice that the situation is
different than for the variable $\sigma_4$; the correlators are indeed
function of such a flat direction whereas they do not depend on
$\sigma_4$. In order to get a sensible result, we therefore need to
include higher order corrections to the two-point functions.

The behavior of $T(\si)$ at small scales is not trivial and many
analytical works have been devoted to its
determination~\cite{Austin:1993rg, Polchinski:2006ee, Dubath:2007mf,
  Copeland:2009dk}. In the Polchinski and Rocha model of
Ref.~\cite{Polchinski:2006ee}, the next-to-leading order terms of the
correlators $\vev{\bXd \cdot \bXd}$ and $\vev{\bXp \cdot \bXp}$ have a
non-integer exponent. These correlators match with Abelian string
simulations performed in Ref.~\cite{Hindmarsh:2008dw} and can also be
used to analytically derive the cosmic string loops distribution
expected in an expanding universe. As shown in
Ref.~\cite{Rocha:2007ni}, these results also match with the scaling
loop distribution observed in the Nambu--Goto numerical simulations of
Ref.~\cite{Ringeval:2005kr}. As a result, we assume in the following a
non-analytical behaviour for $T(\si)$ at small scales
\begin{equation}
\label{eq:NLOcorr}
  T(\s) \simeq {\bar t}^2-\alphac \left(\dfrac{\si}{\corr}\right)^{2\chi}.
\end{equation}
Notice that we are working in the light-cone gauge and therefore leave
$\alphac$ and $\chi$ as undetermined parameters since they cannot be
straightforwardly inferred from the numerics performed in the temporal
gauge. Nevertheless, because the correlation should be smaller as $\s$
becomes larger, $\alphac$ must be positive. Let us also mention the
recent work of Ref.~\cite{Copeland:2009dk} suggesting that at very
small length scales the correlator should become again analytic (with
$\chi=1/2$), i.e. that Eq.~(\ref{eq:NLOcorr}) would hold only for
$\sigma>\sigma_\uc$. However, as discussed in this reference,
$\sigma_\uc$ is shrinking with time in an expanding universe and at
the times of observational interest, Eq.~(\ref{eq:NLOcorr}) is
expected to be valid on all the length scales we are interested in.

With this next-to-leading order form of $T(\s)$, one obtains
\begin{equation}
  \Gamma(\sigma) \simeq \tb^2 \sigma^2 -
  \dfrac{\alphac}{(1+\chi)(1+2\chi)\corr^{2\chi}} \sigma^{2\chi+2},
\end{equation}
and Eq.~(\ref{eq:D2Omega}) reads
\begin{equation}
\begin{aligned}
  \lvev{\calD^2} & = -\frac12 \tb^2 \ka^{ij} \si_{i4} \si_{j4} +
  \alphac \frac{\ka^{ij} \phi(\si_{i4},\si_{j4})
  }{2(2\chi+1)(2\chi+2)\corr^{2\chi}} \,,
\end{aligned}
\end{equation}
where
\begin{equation}
  \phi(\si_{i4},\si_{j4}) \equiv \left( |\si_{i4}|^{2\chi+2}
    +|\si_{j4}|^{2\chi+2} - |\si_{i4}-\si_{j4}|^{2\chi+2} \right).
\end{equation}
We can perform a linear coordinate transformation by introducing the
set of orthonormal unit vectors $(\be_1,\be_2,\be_1\wedge\be_2)$ and
define three new coordinates $\chi_1$, $\chi_2$ and $\chi_3$ along
these directions:
\begin{equation}
\label{eq:trans}
\begin{aligned}
  \chi_1 & \equiv \delta^{ij}(\be_1 \cdot \bk_i ) \si_{j4}, \qquad
  \chi_2 \equiv \delta^{ij} (\be_2 \cdot \bk_i) \si_{j4},\\
  \chi_3 & \equiv \varepsilon^{ijl} (\be_1\cdot \bk_i)(\be_2\cdot \bk_j)
  \si_{l4}.
\end{aligned}
\end{equation}
We then have
\begin{equation}
\begin{aligned}
  \lvev{\calD^2}  = \frac12 \tb^2 (\chi_1^2 + \chi_2^2) & +
  \frac{\alphac}{2(2\chi+1)(2\chi+2)\corr^{2\chi}} \\ & \times \ka^{ij}
  \phi\left[\si_{i4}(\vec{\chi}),\si_{j4}(\vec{\chi})\right] \,.
\end{aligned}
\end{equation}
The third coordinate $\chi_3$ appears in $\vev{\calD^2}$ only when the
next-to-leading order terms in $T(\s)$ are taken into account, which
is consistent with the observation that there is a flat direction at
leading order.  The last term in the previous equation contributes
little to the integrations over $\chi_1$ and $\chi_2$.  Hence we can
safely say that the only non-vanishing component of $\vec{\chi}$ in
the last term is $\chi_3$. This is equivalent to include the next-to-leading order corrections only along the flat direction, i.e. for
\begin{equation}
\label{eq:flatdir}
  \si_{l4} = \frac{1}{\calJ} \varepsilon_{l}^{\phantom{l}ij}
  (\be_1\cdot\bk_i) (\be_2
  \cdot \bk_j) \chi_3,
\end{equation}
where $\calJ$ is the Jacobian of the transformation given by
Eq.~(\ref{eq:trans}).  Then, introducing the outer product coordinates
by
\begin{equation}
\begin{aligned}
w_{ij} & \equiv (\be_1\cdot \bk_i)(\be_2 \cdot \bk_j) - (\be_1\cdot
  \bk_j)(\be_2 \cdot \bk_i) \\
& = \pm \sqrt{k_i^2 k_j^2 - \kappa_{ij}^2}\, ,
\end{aligned}
\end{equation}
one can show that
\begin{equation}
\begin{aligned}
  \ka_{11}\phi(\si_{14},\si_{14}) &= - \frac{2}{\calJ^{2\chi+2}}
  k_1^2 |w_{23}|^{2\chi+2} \chi_3^{2\chi+2}, \\
  \ka_{12}\phi(\si_{14},\si_{24}) &= - \frac{1}{\calJ^{2\chi+2}}
  \ka_{12} \left(|w_{23}|^{2\chi+2} - |w_{34}|^{2\chi+2} \right. \\\ & +
  \left.
    |w_{31}|^{2\chi+2}\right) \chi_3^{2\chi+2},
\end{aligned}
\end{equation}
and other permutations. Finally, making use of identities such as
\begin{equation}
  \ka_{12} + \ka_{13} + \ka_{14}  = \bk_1^2,
\end{equation}
one gets
\begin{equation}
\label{eq:Dsquare}
\begin{aligned}
  \lvev{\calD^2} = \frac12 \tb^2 (\chi_1^2 + \chi_2^2) & +
  \dfrac{\alphac}{(2\chi+1)(2\chi+2)\corr^{2\chi}} \why^2 \\ & \times
  \left( \frac{\chi_3}{\calJ}\right)^{2\chi+2},
\end{aligned}
\end{equation}
with
\begin{eqnarray}
\label{eq:whydef}
\why^2 \equiv -\ka_{12} |w_{34}|^{2\chi+2} + \circlearrowleft\,.
\end{eqnarray}
Notice that $\why^2 \ge 0$ for any quadrilateral because of the
inequality $\vev{\calD^2} \ge 0$. With these new variables, the
trispectrum reads
\begin{equation}
\label{eq:trichi123}
\begin{aligned}
  & T(\bk_1,\bk_2,\bk_3,\bk_4) \simeq \ep^4 \dfrac{L}{4 \Area
    k_1^2k_2^2k_3^2k_4^2} \int \ud\chi_1 \ud \chi_2 \dfrac{\ud
    \chi_3}{\calJ} \\ & \big\{\kappa_{12} \kappa_{34}
    V[\si_{12}(\vec{\chi})]V[\si_{34}(\vec{\chi})] +
    \circlearrowleft \big\}
  e^{-\half \vev{D^2}}.
\end{aligned}
\end{equation}
At this stage, the Gaussian integrations over $\chi_1$ and $\chi_2$
are always finite, and for large enough wave numbers, i.e. $k\corr \gg
1$, we can safely extend the integration range to infinity and also
put $\chi_1=\chi_2=0$ in $V(\sigma_{ij})$. From
Eq.~(\ref{eq:Dsquare}), the integration over $\chi_1$ and $\chi_2$
yields
\begin{equation}
\label{eq:trichi3}
\begin{aligned}
  & T(\bk_1,\bk_2,\bk_3,\bk_4) \simeq \frac{\pi
    \ep^4}{\tb^2}\frac{L}{\Area k_1^2k_2^2k_3^2k_4^2} \int \ud \left( \dfrac{\chi_3}{\calJ} \right) \\
  & \times \big\{ \ka_{12}\ka_{34}
  V\negthinspace\left(w_{34}\dfrac{\chi_3}{\calJ} \right)
  V\negthinspace\left(w_{12}\dfrac{\chi_3}{\calJ} \right) + \circlearrowleft\
  \big\} \\ &\times \exp \left[- \betac \why^2
    \left(\dfrac{\chi_3}{\calJ}\right)^{2 \chi+2}\right],
\end{aligned}
\end{equation}
with
\begin{equation}
  \betac \equiv \dfrac{\alphac}{2 \corr^{2\chi}(2\chi+1)(2\chi+2)}\,.
\end{equation}
The integration over $\chi_3$ may, \emph{a-priori}, be performed in the same
way. However, from Eq.~(\ref{eq:whydef}), one can show that there is
some particular configurations for which $\why$ vanishes
(parallelograms). As a result, one cannot push the integration up to
infinity for those and one has to integrate only over the total string
length. Notice that the integral depends on $L$ only for the
particular parallelogram configurations. As soon as $\why^2\ne 0$, the
small angle limit implies that $Y^2$ is large and the exponential
function takes non-vanishing values only around vanishing
$\chi_3$. For this reason, we separate our analysis in two cases and
first focus on the parallelogram case.

\subsection{Parallelogram configurations $\why^2=0$}

For parallelograms, the two opposite wavevectors forming the
quadrilateral are anti-parallel and $\why^2$ strictly vanishes. Without
loss of generality, we assume $\bk_1+\bk_3=0$ and
$\bk_2+\bk_4=0$. In this case, one has $w_{13}=w_{24}=0$ and we define
\begin{equation}
w=w_{12}=w_{23}=k_1k_2 \sin\theta=k_2k_3\sin\theta.
\end{equation}
The integral in Eq.~(\ref{eq:trichi3}) can be evaluated along the flat
direction $\chi_3/\calJ$, which is given by Eq.~(\ref{eq:flatdir}),
\begin{equation}
\si_{14}=\si_{34}=w \dfrac{\chi_3}{\calJ}, \qquad \si_{24}=0.
\end{equation}
The integration range on $\chi_3/\calJ$ is thus $[-\Lambda,\Lambda]$
where
\begin{equation}
\Lambda = \dfrac{L}{2 |w|}\,.
\end{equation}
From Eq.~(\ref{eq:trichi3}), the trispectrum simplifies to
\begin{equation}
\label{eq:tripara}
\begin{aligned}
 T_0(\bk_1,\bk_2,\bk_3,\bk_4) &= \frac{\pi \ep^4
    \vb^4}{\tb^2}\frac{L^2}{\Area k_1k_2k_3k_4 |w|} \\ & \times
  \left[1 + 2 \cos^2(\theta) \dfrac{1}{L/2} \int_0^{L/2}
    \dfrac{V^2(\sigma)}{\vb^4} \ud \sigma \right] \\ & \simeq
  \frac{\pi \ep^4 \vb^4}{\tb^2}\frac{L^2}{\Area k_1^3k_2^3|\sin\theta|}\,,
\end{aligned}
\end{equation}
where we have neglected the integral in the last line. Since the
correlator $V^2(\sigma)$ is expected to be small at distances larger
than the typical correlation length $\corr$, this integral can be
approximated by
\begin{equation}
  \dfrac{1}{L/2}\int_0^{\corr} \dfrac{V^2(\sigma)}{\vb^4}\ud \sigma =
  \dfrac{2}{L \vb^4} \corr V^2(\sigma_0) \le 2 \dfrac{\corr}{L} \ll 1,
\end{equation}
where we have used the mean value theorem with
$\sigma_0\in[0,\corr]$. Under the scaling transformation $\bk_a \to b
\bk_a$, the parallelogram trispectrum in Eq.~(\ref{eq:tripara}) scales
as
\begin{equation}
\label{eq:ucscaling}
  T_0(b \bk_1,b \bk_2,b \bk_3,b \bk_4)=b^{-6}T_0(\bk_1,\bk_2,\bk_3,\bk_4).
\end{equation}
For parallelograms, it is important to recall that the trispectrum
always gets a contribution from the unconnected part of the four-point
function, which is purely given by a Gaussian distribution,
\begin{equation}
  T_0^{\uuc}(\bk_1,\bk_2,\bk_3,\bk_4)=\Area P(k_1) P(k_2) +\circlearrowleft.
\end{equation}
As shown in Ref.~\cite{Hindmarsh:1993pu}, the power spectrum is given
by
\begin{equation}
P(k) = \sqrt{\pi} \ep^2 \frac{L \vb^2}{\Area \tb k^3}\, ,
\end{equation}
and the unconnected part of the trispectrum also scales as
$b^{-6}$. Therefore the non-Gaussian contributions for parallelogram
configurations remain of the same order of magnitude as the Gaussian
ones, with the exception of the squeezed limit $\theta \to 0$. As we
will see in the following, all other quadrilateral configurations have
a scaling law which is different than Eq.~(\ref{eq:ucscaling}).

\subsection{Quadrilateral configurations with $\why^2\gg 1$}

In this case, the integrand in Eq.~(\ref{eq:trichi3}) takes non
vanishing values only around $\chi_3=0$ and we can safely extend the
integration range over $\chi_3/\calJ$ to infinity, as we have done for
$\chi_1$ and $\chi_2$. One gets
\begin{equation}
  \label{eq:triinfty}
\begin{aligned}
  T_\infty(\bk_1,\bk_2,\bk_3,\bk_4) &\simeq
  \ep^4 \frac{\vb^4}{\tb^2}\frac{L \corr}{\Area} \left(\alphac
    \corr^2\right)^{-1/(2\chi+2)} f(\chi) \\ & \times
  g(\bk_1,\bk_2,\bk_3,\bk_4).
\end{aligned}
\end{equation}
The function $f(\chi)$ is a number depending only on the power-law
exponent $\chi$
\begin{equation}
 f(\chi) = \frac{\pi}{\chi+1}\Gamma\left ( \dfrac{1}{2\chi+2} \right) \left[ 4(2\chi+1)(\chi+1)\right]^{1/(2\chi+2)},
\end{equation}
and $g(\{\bk_a\})$ is the trispectrum geometrical factor
\begin{equation}
\label{eq:geometrical}
\begin{aligned}
&  g(\bk_1,\bk_2,\bk_3,\bk_4)  = \dfrac{\ka_{12} \ka_{34} + \ka_{13}
    \ka_{24}+ \ka_{14} \ka_{23}}{k_1^2 k_2^2 k_3^2 k_4^2} \\ & \times 
\left [-\ka_{12} \left(k_3^2 k_4^2 - \kappa_{34}^2\right)^{\chi+1} +
  \circlearrowleft \right] ^{-1/(2\chi+2)}.
\end{aligned}
\end{equation}
From Eq.~(\ref{eq:triinfty}), we can derive various consequences worth
mentioning.  First, the sign of the trispectrum is completely
determined by the geometrical factor in Eq.~(\ref{eq:geometrical}),
which is manifestly symmetric under the permutation of two different
wavevectors. The factor
\begin{eqnarray}
\label{eq:geodenom}
  -\ka_{12} \left(k_3^2 k_4^2 - \kappa_{34}^2\right)^{\chi+1} +
  \circlearrowleft
\end{eqnarray}
is always positive or zero. Also $f(\chi)$ (for a physically
reasonable range of $\chi$) and $\alphac$ are positive. Therefore the
sign of the trispectrum is given by the factor
\begin{eqnarray}
\ka_{12}\ka_{34}+ \ka_{13}\ka_{24}+ \ka_{14}\ka_{23},
\end{eqnarray}
which can be positive or negative according to the quadrilateral under
scrutiny. 

Secondly, under the scaling transformation $\bk_a \to b \bk_a$, the
geometric factor scales as
\begin{equation}
g(b\bk_1,b\bk_2,b\bk_3,b\bk_4)=b^{-\rho} g(\bk_1,\bk_2,\bk_3,\bk_4),
\end{equation}
with
\begin{equation}
\rho = 6 + \frac{1}{\chi+1}\,.
\end{equation}
Contrary to the case of the power spectrum, of the bispectrum, and of
the parallelogram configurations, the leading term of the trispectrum
scales with a non-integer power-law exponent. For $\chi>0$, the
trispectrum decays slightly faster at small scales than the
bispectrum. Let us recap that in the temporal gauge, the string
tangent vector correlation function exponent $\chi$ is a small
quantity related to the expansion rate of the scale factor and to the
mean square velocity of strings~\cite{Polchinski:2006ee}. This is
certainly also the case in the light-cone gauge and one may be able to
use the trispectrum to distinguish between different models of
strings. For instance, in Abelian Higgs numerical simulations, the
strong back-reaction induced by scalar and gauge radiation produces a
mean square velocity lower than in classical Nambu-Goto
simulations~\cite{Bennett:1990, Hindmarsh:2008dw}. Meanwhile, the loop
distribution observed in Nambu--Goto simulations has a power-law
exponent which is uniquely given by
$\chi$~\cite{Ringeval:2005kr}. Interestingly, the scaling exponent is
different from the one associated with parallelogram
configurations. These two different scaling behaviors may actually be
used to distinguish the trispectrum by cosmic string with the one
generated by other sources.

\subsection{Interpolating trispectrum for all quadrilaterals}

When $\why^2\simeq 0$ but non-vanishing, i.e. for quadrilaterals close
to parallelograms, one cannot push the integration range in
Eq.~(\ref{eq:trichi3}) to infinity. Contrary to the case $\why^2=0$,
the integration over $\chi_3/\calJ$ cannot be performed explicitly in
this case. Nevertheless, we can make some approximations. First, for
configurations close to parallelograms, two of the $w_{ij}$ quantities
are expected to be small, say $w_{13}$ and $w_{24}$. For those, one
can replace the $V(\sigma)$ functions in Eq.~(\ref{eq:trichi3}) by
$\vb^2$. On the other hand, one expects the other $w_{mn}$ factors to
be large and Eq.~(\ref{eq:trichi3}) has terms involving the product
$V(\sigma) V(\sigma')$. As for the parallelograms, we expect those to
be at most of the order $\vb^4 \corr/L$, which can be neglected
compared to the terms in $\vb^4$. With an integration range over
$\chi_3/\calJ$ given by $[-\Lambda,\Lambda]$, where
$\Lambda({\bk_a},L)$ has still to be specified, performing the last
integration over $\chi_3$ yields
\begin{equation}
  \label{eq:unsymtri}
\begin{aligned}
  & T_{w_{13}w_{24}}(\bk_1,\bk_2,\bk_3,\bk_4) \simeq \ep^4
  \frac{\vb^4}{\tb^2} \frac{L \corr }{\Area} \left(\alphac
    \corr^2\right)^{-1/(2\chi+2)}f(\chi) \\ & \times
  \gammain{\frac{1}{2\chi+2}}{\betac \why^2 \chimax^{2\chi+2}}
  \dfrac{\kappa_{13} \kappa_{24}}{k_1^2 k_2^2 k_3^2 k_4^2}
  Y^{-2/(2\chi+2)},
\end{aligned}
\end{equation}
where $\gammain{a}{x}$ denotes the normalised incomplete lower gamma
function
\begin{equation}
\gammain{a}{x} \equiv \dfrac{\gamma(a,x)}{\Gamma(a)}\,.
\end{equation}
In the limit $\why^2\to 0$, this expression matches with
Eq.~(\ref{eq:tripara}) for $\Lambda=L/(2 |w_{12}|)$. In order to
interpolate between Eqs.~(\ref{eq:tripara}) and (\ref{eq:triinfty}) we
can replace the geometrical factor in Eq.~(\ref{eq:unsymtri}) by the
factor $g(\{\bk_a\})$ and chose the cutoff $\Lambda$ to be
\begin{equation}
\label{eq:lambda}
\begin{aligned}
  \Lambda & \equiv 
  \dfrac{2 L}{|w_{12}|+|w_{13}|+|w_{14}|
    + |w_{23}| + |w_{24}| + |w_{34}|} \\ & \times \dfrac{k_1 k_2 k_3
    k_4}{\kappa_{12}\kappa_{34}+\kappa_{13}\kappa_{24} +
    \kappa_{14}\kappa_{23}} \,.
\end{aligned}
\end{equation}
Our interpolation formula for the trispectrum finally reads
\begin{equation}
  \label{eq:finaltri}
\begin{aligned}
  & T(\bk_1,\bk_2,\bk_3,\bk_4) \simeq \ep^4 \frac{\vb^4}{\tb^2}
  \frac{L \corr }{\Area} \left(\alphac \corr^2\right)^{-1/(2\chi+2)}f(\chi)
  \\ & \times \gammain{\frac{1}{2\chi+2}}{\betac
      \why^2 \chimax^{2\chi+2}} g(\bk_1,\bk_2,\bk_3,\bk_4) ,
\end{aligned}
\end{equation}
with $\Lambda$ given by Eq.~(\ref{eq:lambda}) and $g(\{\bk_a\})$ by
Eq.~(\ref{eq:geometrical}). For $\why^2$ large, the gamma function is
close to one and we recover Eq.~(\ref{eq:triinfty}). The limit
$\why^2=0$ gives again the leading order of Eq.~(\ref{eq:tripara}).

\section{Geometrical factors of symmetric quadrilaterals}

\label{sec:quad}

In this section, we explore the dependency of the trispectrum
geometrical factor given by Eq.~(\ref{eq:geometrical}) for some
symmetric quadrilateral configurations of the wavevectors.

\begin{figure}
\begin{center}
\includegraphics[width=8.5cm]{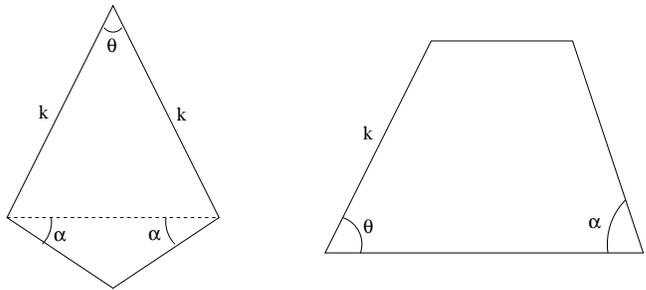}
\caption{Quadrilateral configurations for the trispectrum
  wavenumbers. The left is referred to as the ``kite'' quadrilaterals
  with two adjacent sides of equal length and the other two sides also
  of equal length. The right quadrilateral is a ``trapezium'' which is
  defined to have two opposite sides parallel.}
\label{fig:quad}
\end{center}
\end{figure}

\subsection{Kite configurations}

Let us first consider a quadrilateral like the one given in
Fig.~\ref{fig:quad} (left). From Eq.~(\ref{eq:whydef}), one gets
\begin{equation}
\label{eq:y2y}
\begin{aligned}
\why^2 = k^{6+4\chi} y^2(\theta,\alpha),
\end{aligned}
\end{equation}
with
\begin{equation}
\begin{aligned}
&  y^2(\theta,\alpha)  = \left[\sin^2(\theta/2)\right]^{1+\chi} \\ & \times
  \Bigg\{2 \sin(\theta/2) \dfrac{\sin(\alpha-\theta/2)}{\cos \alpha}
  \times \left[\dfrac{\cos^2(\alpha - \theta/2)}{\cos^2 \alpha}
  \right]^{1+\chi} \\ & - 2 \sin(\theta/2)
  \dfrac{\sin(\alpha+\theta/2)}{\cos \alpha}
  \left[\dfrac{\cos^2(\alpha + \theta/2)}{\cos^2 \alpha}
  \right]^{1+\chi} \\ &+ 4^{1+\chi} \sin^2(\theta/2)
  \left[\cos^2(\theta/2)\right]^{1+\chi}
  \dfrac{\cos(2\alpha)}{\cos^2(\alpha)} \\ & - 4^{1+\chi} \cos(\theta)
  \left[\sin^2(\theta/2) \tan^2(\alpha) \right]^{1+\chi} \Bigg\}.
\end{aligned}
\end{equation}
From Eq.~(\ref{eq:geometrical}), the geometrical factor reads
\begin{equation}
\label{eq:kiteexact}
\begin{aligned}
  g(\bk_1,\bk_2,\bk_3,\bk_4) & = \dfrac{\cos^2(\alpha)\left[1-
      2\cos(2\alpha)\cos(\theta) \right]}{\sin^2(\theta/2)} \\ & \times
  \dfrac{1}{k^{\rho} y^{2/(2+2\chi)}}\,.
  \end{aligned}
\end{equation}
As expected from the trispectrum scaling law, the kite trispectrum
decays as $1/k^\rho$ at small angular scales. The overall amplitude is
however amplified for squeezed configurations and diverges for $\theta
\to 0$. For $\theta$ small, the leading terms of the previous
expression are
\begin{equation}
  \label{eq:kitesq}
\begin{aligned}
  g & \underset{\theta \ll 1}{\sim} \dfrac{8 \cos^2(\alpha)}{k^\rho \theta^{\rho-3}} (1-2 \cos
  2\alpha ) \\ & \times \bigg \{ 2(1+\chi) \tan^2(\alpha) - 1 + 4^\chi
  (1-\tan^2 \alpha) \bigg \}^{-1/(2\chi+2)}.
\end{aligned}
\end{equation}
The sign of the kite trispectrum is the same as $1-2\cos(2\alpha)
\cos(\theta)$ and, at small $\theta$, is negative for $\alpha<\pi/6$
and positive otherwise. As for the bispectrum, we recover that
squeezed configurations are the most sensitive to a string signal,
certainly due to the elongated temperature discontinuities induced by
the GKS effect.
\begin{figure}
\begin{center}
\includegraphics[width=8.5cm]{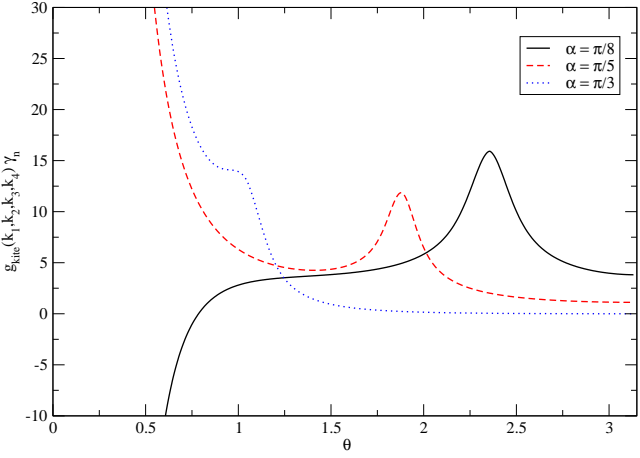}
\caption{Trispectrum geometrical factor for the kite quadrilaterals as
  a function of $\theta$, plotted for various values of $\alpha$. The
  trispectrum is enhanced in the squeezed limit $\theta \to 0$. The
  bump for $\theta_\up=\pi-2\alpha$ corresponds to the parallelogram
  limit for which the unconnected part is no longer vanishing.}
\label{fig:kite}
\end{center}
\end{figure}
In Fig.~\ref{fig:kite}, we have represented the full
geometrical dependency coming from Eq.~(\ref{eq:finaltri}) as a
function of $\theta$ and for various values of $\alpha$. For
convenience, we have chosen $\chi=0.29$, $\betac=1$, $k=1$ and
$L=20$. The incomplete gamma function contributes for configurations
close to the parallelogram ones which appear as a bump in
Fig.~\ref{fig:kite} for $\theta_\up=\pi-2\alpha$. For the kites, the
argument of the gamma function simplifies to
\begin{equation}
\begin{aligned}
   \betac \why^2 \chimax^{2\chi+2} & = k^2 \dfrac{\alphac
    \corr^2}{2(2\chi+1)(2\chi+2)}
  \left(\dfrac{2L}{\corr}\right)^{2\chi+2}y^2(\theta,\alpha) \\ &
  \times \dfrac{\left[ 1-2\cos(2\alpha) \cos(\theta)
    \right]^{-2(\chi+1)}} {\left\{ 2 \sin(\theta) +
      \left[\cos(\theta)-1\right]\tan(\alpha) \right\}^{2(\chi+1)}}
  \,.
\end{aligned}
\end{equation}
As can be seen on this plot, we recover the change of sign when
$\alpha$ crosses the value $\pi/6$. The bump at
$\theta_\up=\pi-2\alpha$ corresponds to the parallelogram limit of the
kite configuration for which $y^2(\theta,\alpha)\to 0$.

\subsection{Trapezium}

Let us next consider a quadrilateral given by the right side of
Fig.\ref{fig:quad} having two opposite sides parallel. Without lost of
generality, one can assume that the upper side is of smaller length
than the bottom. Denoting their ratio by $\sin^2(\beta)$, after some
algebra, the factor $\why^2$ is still given by Eq.~(\ref{eq:y2y}) with
\begin{equation}
\begin{aligned}
  & y^2(\theta,\alpha) = \left[\sin^2(\theta)\right]^{\chi+1} \left[
    \dfrac{\sin^2(\alpha+\theta)}{\sin^2(\alpha)} \right]^{\chi+2} \\ &
    \times \dfrac{1 - \left[\cos^2(\beta) \right]^{2\chi+1} -
      \left[\sin^2(\beta) \right]^{2\chi+1}}{\tan^2(\beta)
      \left[\sin^2(\beta)\right]^{2\chi+2}}\,. 
\end{aligned}
\end{equation}
Similarly, the geometrical factor reads
\begin{equation}
\label{eq:geotrap}
\begin{aligned}
  & g(\bk_1,\bk_2,\bk_3,\bk_4) = \dfrac{\sin(\alpha) \sin(\theta) - 3
    \cos(\alpha) \cos(\theta)}{k^{\rho}} \\ & \times
  \dfrac{\sin(\alpha)}{\sin^2(\theta)}
  \left[\dfrac{\sin^2(\alpha)}{\sin^2(\alpha+\theta)}
  \right]^{(\rho-3)/2} \sin^4(\beta)
  \left[\tan^2(\beta)\right]^{(\rho-4)/2} \\ & \times \left\{1-
    \left[\cos^2(\beta) \right]^{2\chi+1} - \left[ \sin^2(\beta)
    \right]^{2 \chi+1} \right\}^{-1/(2\chi+2)}.
\end{aligned}
\end{equation}
As expected, the trapezium trispectrum decays with the power-law
exponent $k^{-\rho}$. The overall amplitude is again amplified for
elongated configurations and diverges for $\theta\to 0$. For convex
quadrilaterals, assuming $0<\theta<\pi-\alpha$, the sign of the
trispectrum is given by the first term of Eq.~(\ref{eq:geotrap}). As a
result, it is is negative for $\theta<\theta_\us$ and positive
otherwise, where $\theta_\us$ is given by
\begin{equation}
  \theta_\us = \arccos\left[\dfrac{\sin(\alpha)} {\sqrt{9 \cos^2(\alpha) + \sin^2(\alpha)}}\right].
\end{equation}
For an isosceles trapezium with $\alpha=\theta$, the change of sign
occurs at $\theta_\us=\pi/3$.
\begin{figure}
\begin{center}
\includegraphics[width=8.5cm]{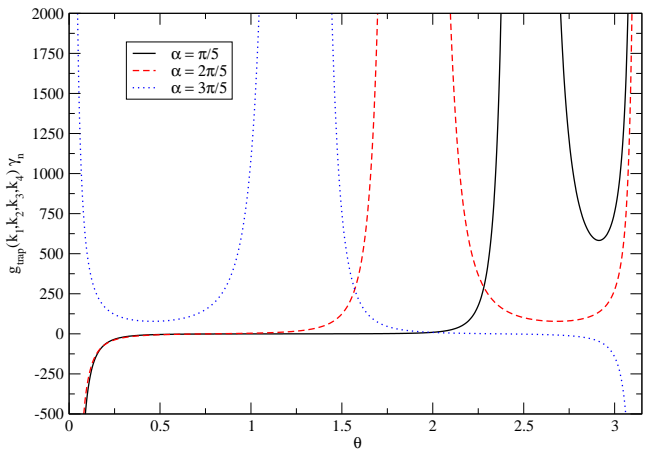}
\caption{Trispectrum geometrical factor for the trapezium
  quadrilaterals as a function of $\theta$, plotted for various values
  of $\alpha$. For convenience, the ratio of the two parallel sides
  has been fixed to $3/4$ and $\chi=0.29$. The divergence in the
  squeezed limit occurs at $\theta \to 0$ but also at
  $\theta_\up=\pi-\alpha$ for infinitely elongated parallelograms. For
  $\theta>\theta_\up$, the trapeziums are no longer convex and
  represent ``butterfly'' configurations which are squeezed for
  $\theta \to \pi$.}
\label{fig:trap}
\end{center}
\end{figure}
Finally, in Fig.~\ref{fig:trap}, we have plotted the full geometrical
dependence as a function of $\theta$, for various values of
$\alpha$. For the trapeziums, the argument of the gamma function is
\begin{equation}
\begin{aligned}
  & \betac \why^2 \chimax^{2\chi+2} = k^2 \dfrac{\alphac
    \corr^2}{2(2\chi+1)(2\chi+2)}
  \left(\dfrac{2L}{\corr}\right)^{2\chi+2} y^2(\theta,\alpha) \\ &
  \times
  \left[\sin(\alpha)\sin(\theta)-3\cos(\alpha)\cos(\theta)\right]^{-2(\chi+1)}
  \\ & \times \left\{\dfrac{\sin(\theta)\sin(\theta+\alpha)
      \left[3+\cos^2(\beta)\right]} {\sin(\alpha)
      \sin^2(\beta)}\right\}^{-2(\chi+1)} .
\end{aligned}
\end{equation}
The divergence for the parallelograms visible at $\theta=\pi-\alpha$
comes again from the squeezed shape. Imposing a fixed value of
$\sin^2(\beta)$ implies that such parallelograms are infinitely
elongated. The configuration with $\theta>\pi-\alpha$ are
self-intersecting trapeziums having a butterfly shape. Their squeezed
limit occurs for $\theta\to\pi$ for which the trispectrum is again
strongly enhanced.

\section{Conclusion}

\label{sec:concl}

In this paper, we have analytically derived the CMB temperature
trispectrum induced by cosmic strings using the string correlation
functions in the Gaussian approximation. The trispectrum generically
decays with a non-integer power-law behaviour at small angular scales
which depends on the string microstructure through the behaviour of
the tangent vector correlator on small distances. Its eventual
detection and measurement may therefore help to distinguish between
different string models. We have also found that the trispectrum
diverges, in the framework of our approximations, on all squeezed
configurations whose measurements remain however limited by the finite
experimental resolution. In fact, such a non-integer power-law is
linked to the existence of a ``flat direction'' at leading order and
the four-point function ends up being sensitive to the next-to-leading
order string tangent vector correlator. This situation is also present
in the n-point function and we do expect all of the higher n-point
function to exhibit non-integer power-law behaviours. Since this
situation was not encountered for the two- and three-point functions,
the next step will be to compare our results here with the trispectrum
computed from CMB maps obtained by string network simulations.

Finally, let us notice that we have not attempted to make any
comparison with a CMB trispectrum produced by primordial
non-Gaussianities of inflationary origin. The situation is nearly the
same as it is for the string bispectrum~\cite{Hindmarsh:2009qk}. The
so-called $\tauNL$ and $\gNL$ parameters quantify the amplitude of the
primordial four-point function of the curvature perturbation on
super-Hubble scales. As a result, the induced trispectrum of the CMB
temperature fluctuations strongly depends on the CMB transfer
functions and exhibits damped oscillations with respect to the
multipole moments. Here, we have direcly derived the CMB temperature
trispectrum produced by the strings and it would therefore make no
sense to find an associated $\tauNL$ and $\gNL$. An alternative
approach might be to estimate what values $\tauNL$ and $\gNL$ would
assume in a primordial-type oriented data analysis if the
non-Gaussianities were actually due to strings. This could be done
with a Fisher matrix analysis for a given experiment but we leave this
question for a forthcoming work.

\acknowledgments
This work is partially supported by the Belgian Federal Office for
Scientific, Technical, and Cultural Affairs through the
Inter-University Attraction Pole Grant No. P6/11.

\bibliography{bibstrings}
\end{document}